\documentclass[9pt,twocolumn,twoside]{osajnl}

\journal{ol} 

\setboolean{shortarticle}{true}
\usepackage{graphicx}
\makeatletter

\newcommand{\Rmnum}[1]{\expandafter\@slowromancap\romannumeral #1@}
\makeatother

\title{$\mathcal{PT}$-symmetric topological near-zero interface state}

\author[1]{Zhenjuan Liu}
\author[1]{Kaiwen Ji}
\author[1]{Haohao Wang}
\author[1]{Yanan Dai}
\author[2]{Yuanmei Gao}
\author[3]{Yishan Wang}
\author[1*]{Xinyuan Qi}
\author[1]{Jintao Bai}
\affil[1]{State Key Laboratory of Photoelectric Technology and Functional Materials, International Joint Research Center on Photoelectric Technology and Functional Nanomaterials, School of Physics, Northwest University, 710127 Xi'an, China}
\affil[2]{Shandong Provincial Key Laboratory of Optics and Photonic Devices, College of Physics and Electronics, Shandong Normal University, Jinan, 250014, China}
\affil[3]{State Key Laboratory of Transient Optics and Photonics, Chinese Academy of Sciences, Xi'an 71068, China}
\affil[*]{Corresponding author: qixycn@nwu.edu.cn}

\begin{abstract}
Photonic systems with parity-time (PT) symmetry and topology are attracting considerable attentions. In this work, topological near-zero interface state is studied in cleverly designed qusi-one-dimensional $\mathcal{PT}$-symmetric photonic lattice. Further study shows that such topological interface state experiences phase transition like bulk states and thus supports real-eigenvalued topological states in $\mathcal{PT}$-symmetric system, which finally resuts in stable topological interface state in  $\mathcal{PT}$-symmetric quasi-one dimensional photonic lattice. Our study enriches the content of non-Hermitian topological physics and might have potential applications in the fields of topological lasing and  quantum computation. 
\end{abstract}

\setboolean{displaycopyright}{true}

\usepackage{CJKutf8}
\usepackage{amssymb}
 
\begin{document}
\begin{CJK*}{UTF8}{gbsn}

\maketitle

The zero-energy particle, also referred to as zero mode with its eigen energy pinned at the middle of a gapped band structure, such as Majorana fermion, has attracted considerable attention due to its important role in fault-tolerant quantum computation~\cite{Moore1991Nonabelions,Fu2008Superconducting,Levene2003Zero}. Topological zero edge modes in one dimensional chiral system, which are immune to the perturbations and continuous changes, have been realized in photonic superlattice~\cite{2009Observation}. Since then, the Su-Schrieffer-Heeger (SSH)  model or related models have been discussed and realized in many systems, such as photonic crystals, electromagnetic metamaterials, plasmonic waveguide arrays, polariton micropillars and coupled optical waveguides~\cite{2014Photonic,2014Topological,2015Kibble,saei_ghareh_naz_topological_2018}. The study of topological physics with photons allows for the exploration of phenomena inaccessible in the context of condensed matter. 

A case in point is non-Hermiticity in the form of optical gain and loss. In photonics, gain and loss is much more common than in electrons in solids: gain media are the basis for lasers, and loss of photons is ubiquitous in every photonic device. A series of works have been delved into the optical effects with both non-Hermiticity and topology~\cite{2018Topological,Lieu2017Topological}. Inspired by the model proposed in References~\cite{2009Topological,zeuner_observation_2015}, optical waveguide arrays were employed to demonstrate that the winding number of a one-dimensional topological system could be extracted from a non-Hermitian quantum walk. Photonic “tachyonlike” dispersion~\cite{2011PT} demonstrated in the form of exceptional rings in photonic crystals~\cite{zhen_spawning_2015}, “Fermi arc”-type states that connect between exceptional points in two-dimensional systems~\cite{2017Observation} as well as optical funnel, an optical effects that all the optical modes will be trapped in a certain site, has been realized in non-Hermitian topological mesh~\cite{Weidemann311}.

Generally speaking, the theory of Hermitian topological photonics can not be extended directly to non-Hermitian topological system. As a special case, since the Berry phase of a parity-time ($\mathcal{PT}$)-symmetric topological  system is only decided by the counterpart in Hermtian system~\cite{Weimann2016Topologically,Zhao2016Robust}, $\mathcal{PT}$-symmetric systems have been a natural platform to study non-Hermitian topological effects. Refs.~\cite{2011Edge,2011Absence} report that spontaneous breaking edge states occur when the topological system is $\mathcal{PT}$-symmetric. This means such a system supports complex eigenvalues, which results in unstable light propagation even for the topological edge state. In 2016, topologically protected bound states at the interface was achieved in one-dimensional modulated photonic lattice with $\mathcal{PT}$-symmetry~\cite{Weimann2016Topologically}. This work provides us a novel view to realize the stable topological state in non-Hermitian $\mathcal{PT}$-symmetric topological system. However, in one dimensional(1D) system, topological phases of matter cannot exist without imposing strict symmetry conditions, such as chrial symmetry or particle-hole symmetry~\cite{2009Periodic,2009Classification,Weimann2016Topologically}.  Recent work from our group~\cite{Kaiwen2019Asymmetric} demonstrates that near-zero edge modes may survive in quasi-1D photonic lattice without chiral or particle-hole symmetries. However, systematic research on the influence of non-Hermitian terms on the topological near-zero modes in $\mathcal{PT}$-symmetric system is still absent and corresponding light dynamics in such a systems hasn't yet been realized.  

In this Letter, a quasi-1D topological photonic lattice with $\mathcal{PT}$ symmetry is theoretically constructed and the systematical analysis of the impact for the  non-Hermitian terms on the energy spectrum  is shown. Significantly, we realize a stable topological near-zero interface mode in a quasi-1D binary photonic lattice with global $\mathcal{PT}$ symmetry. The numerical simulation agrees well with the theoretical analysis. Our study paves a new way to the realization of stable topological states in $\mathcal{PT}$-symmetric photonic system  and might have potential applications in the fields of topological photonics. 

\begin{figure}
\centering
\fbox{\includegraphics[width=\linewidth]{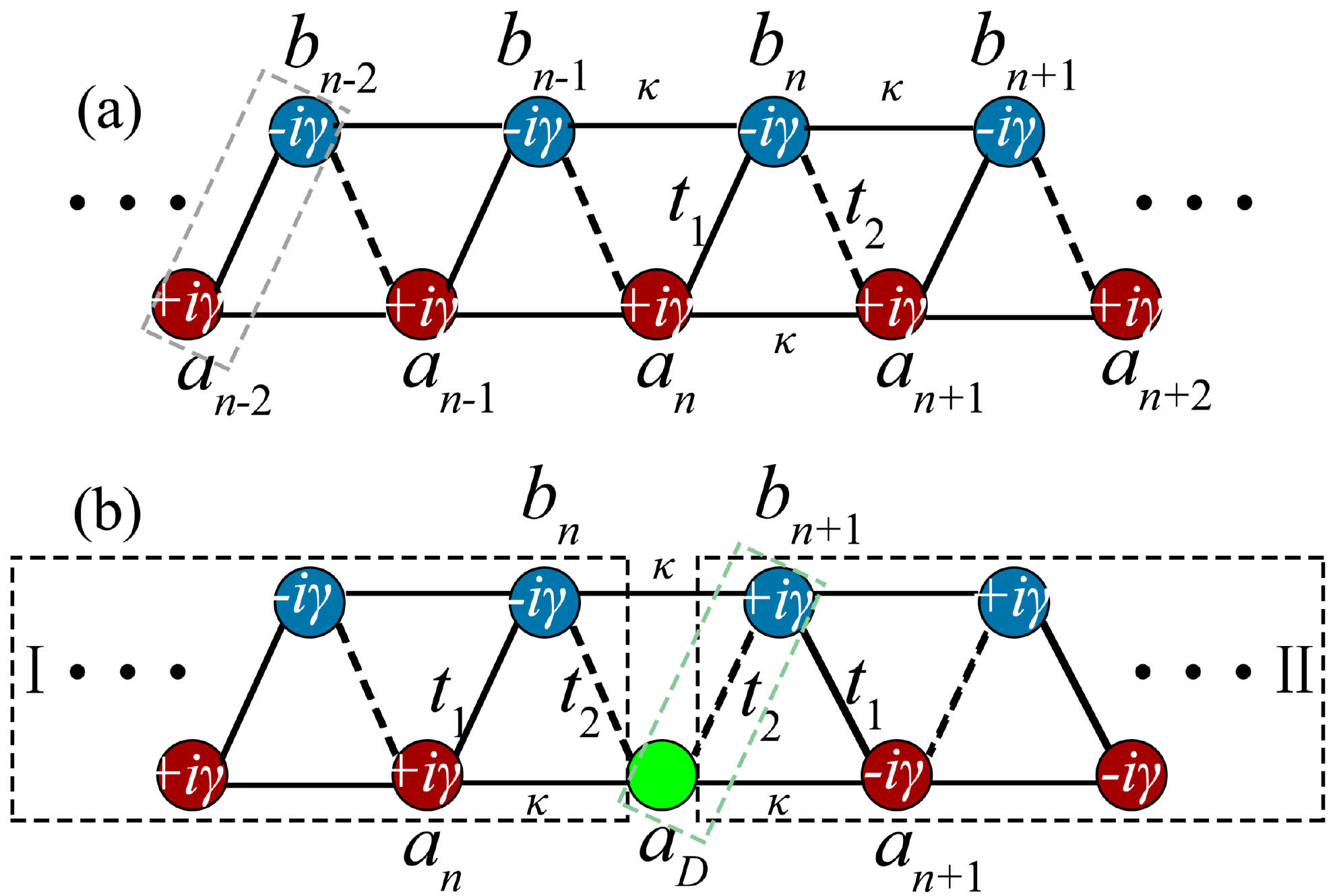}}
\caption{Structures of two-layered photonic lattice. (a), $\mathcal{PT}$-symmetric waveguide array with coupling constants $t_1$(interlayer solid lines), $t_2$(interlayer dashed lines) and $\kappa$(intralayer solid lines) as well as alternating onsite potential $i\gamma$ and $-i \gamma$ (describing gain and loss in  photonic applications, respectively). Solid red and blue dots, waveguides with field amplitudes of $a_n$ and $b_n$, respectively. (b), Binary $\mathcal{PT}$-symmetric photonic lattice with the introduction of the interface waveguide, $a_D$. Dashed boxes in two subfigures represent a unit cell.}
\label{fig.1}
\end{figure}

We first construct a two-layered photonic lattice and introduce alternating gain and loss in a $\mathcal{PT}$-invariant fashion to extend the system to be non-Hermitian, as shown in Fig.~\ref{fig.1}(a). In coupled-mode theory, our single-mode-coupled waveguide lattice can be modeled by a tight-binding system as follows

\begin{equation} \tag{1a}
\begin{aligned}
i \frac{d}{dz} \psi_n^{(a)} = i \gamma \psi_n^{(a)}+t_1 \psi_n^{(b)}+t_2 \psi_{n-1}^{(b)}+\kappa (\psi_{n-1}^{(a)}+\psi_{n+1}^{(a)}),\\
\end{aligned}
\label{eq:refname1a}
\end{equation}

\begin{equation} \tag{1b}
\begin{aligned}
i \frac{d}{dz} \psi_n^{(b)} = -i \gamma \psi_n^{(b)}+t_1 \psi_n^{(a)}+t_2 \psi_{n+1}^{(a)}+\kappa (\psi_{n-1}^{(b)}+\psi_{n+1}^{(b)}).
\end{aligned}
\label{eq:refname1b}
\end{equation}
where $\psi_n^{(a)}$ and $\psi_n^{(b)}$ are the field envelopes for waveguides $a_n$ and $b_n$ in each unit cell,  $t_1$, $t_2$ and $\kappa$ denote the coupling constants. $\gamma$ is the magnitude of gain/loss and $n$ denotes the unit cell number ranging from 0 to $N$. The Hamiltonian of such a system can be written as

\begin{equation}\tag{2}
\begin{aligned}
\mathcal{H}=h_0\delta_0+h_x\delta_x+h_y\delta_y+i\gamma\delta_z,
\end{aligned}
\label{eq:refname2}
\end{equation}

\begin{equation}\tag{3}
\begin{aligned}
h_0=2\kappa cosk, \qquad h_x=t_1+t_2 cosk, \qquad h_y=t_2 sink.
\end{aligned}
\label{eq:refname3}
\end{equation}
where $k$ is the wave number in the Bloch zone, $\delta_0$ is the identity matrix and $\delta_{x,y,z}$ are the Pauli matrices. The energy dispersion and eigen vectors of the Hamiltonian can be calculated with these parameters, which read

\begin{equation}\tag{4}
E_\pm=\pm \sqrt{h_x^2+h_y^2-\gamma^2}+h_0,
\label{eq:refname4}
\end{equation}

\begin{equation}\tag{5}
|\psi_\pm \rangle=( \frac{i\gamma\pm \sqrt{h_x^2+h_y^2-\gamma^2}}{h_x+i h_y},1 )^T.
\label{eq:refname5}
\end{equation}

In Hermitian systems, Winding number and Chern number are often analyzed to determine the topological invariants. However, such parameters cannot be well defined in non-Hermitian systems. To characterize the topological nature in the non-Hermitian system, the global Berry phase $\phi_B$ ~\cite{Weimann2016Topologically,Zhao2016Robust} is used which corresponds to the summation of complex Berry phases $\phi_B^{-}$ in the lower and $\phi_B^{+}$ in the upper bands in our case. The Berry phase in each band can be calculated: $\phi_B^{\pm}=\oint_k A_\pm dk$, where $A_{\pm}=i \langle\psi_{\pm}| \partial_k |\psi_{\pm} \rangle$ is the Berry connection, and $\langle\psi_{\pm}|$ and $|\psi_{\pm} \rangle$ are the normalized left and right eigenvectors of the Hamiltonian $\mathcal{H}$, These two complex Berry phases follow as
\begin{equation}\tag{6}
\phi_B^\pm=\frac{\phi_0}{2}\pm\frac{1}{2} \oint_{\phi_k} cos\gamma_k d\phi_k ,
\label{eq:refname6}
\end{equation}
\begin{figure}
\centering
\fbox{\includegraphics[width=\linewidth]{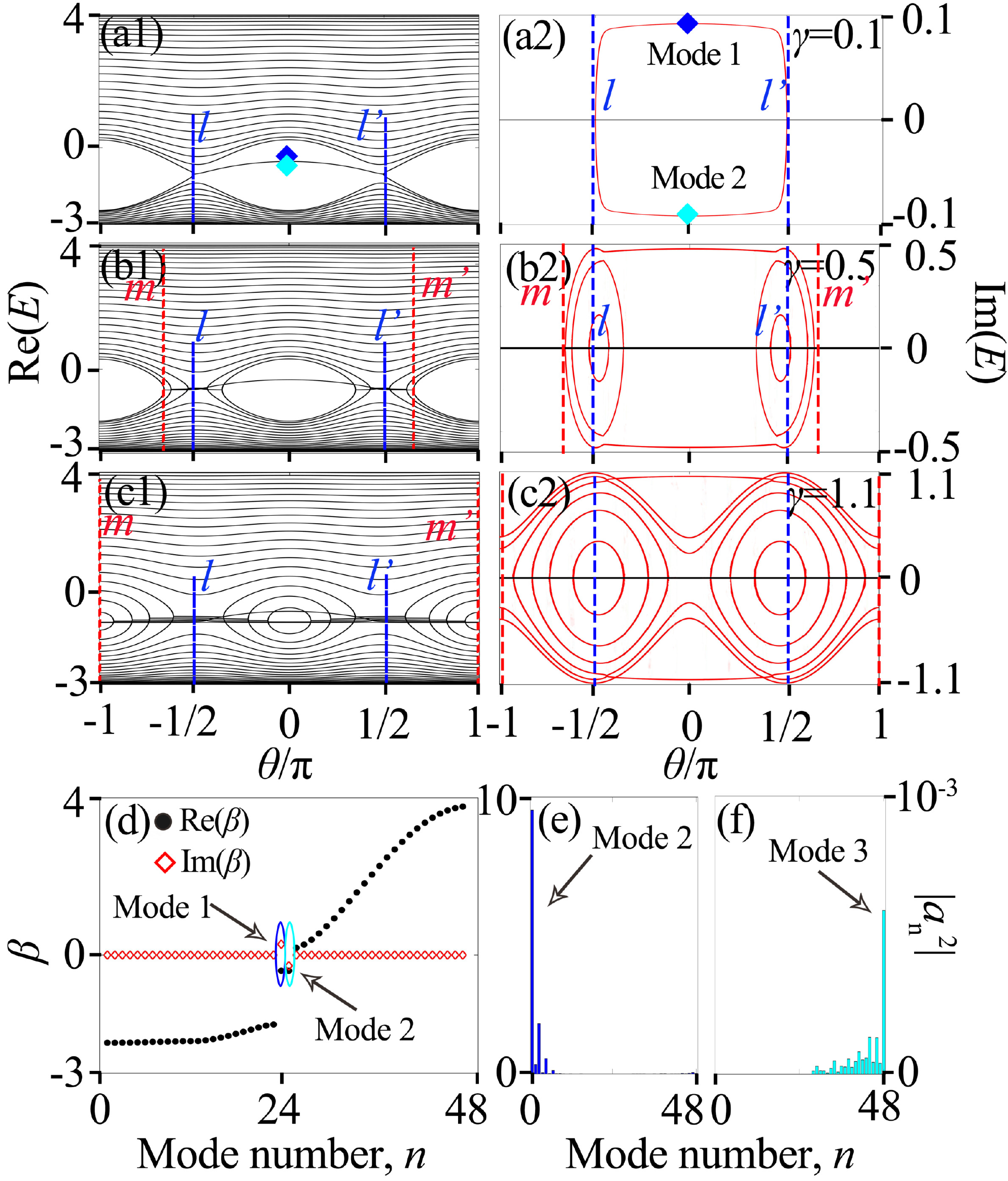}}
\caption{Eigenvalue spectrum of the two-layered photonic lattice as a function of $\theta$ with parameters $\delta = 1/3$, $t = 3/2$ and $\kappa=0.2$ for different $\gamma$: (a1) $\gamma = 0.1$, (b1) $\gamma=0.5$ and (c1) $\gamma=1.1$ under the OBC ($N_a$=$N_b$=24). The region denoted by the blue dashed lines $l$ and $l'$ represents the topological nontrivial regime ($\theta \in [-\pi/2,\pi/2]$) while the red dashed lines \textit{m} and \textit{$m'$} mark the degenrated point in topological trivial regime. (d), Propagation constants and intensity distributions for Mode 1 (e) and 2 (f), respectively. The numbers of waveguides are $N_a$=$N_b$=24. Other parameters are  \textit{t}=3/2, $\delta$=1/3, $\kappa=0.2$ and $\theta=0$.}
\label{fig.2}
\end{figure}
where $\gamma_k=\arctan(\rho_k/i\gamma)$, $\rho_k=|t_1+t_2 exp(-ik)|$ and $\phi_k=\arg[t_1+t_2 exp(-ik)]$. Clearly, the global Berry phase $\phi_B$ remains quantized independent of onsite gain/loss in the system (i.e., $\phi_B^+ + \phi_B^-=\phi_0$) thus the existence of edge states in our system won't be influenced by the introduction of gain or loss.  Here, in Hermitian case($\gamma=0$), the Zak phase can be learned from the relation $\phi_0^{\pm}=i \int \limits_0^{2\pi} dk \langle \psi_{\pm}| \partial_k |\psi_{\pm} \rangle= 2 \pi W$. W takes on the value 1 when there are edge states and 0 when there are not. One can deduce that $W$=1 when $0<t_1/t_2<1$ while $W$=0 when $t_1/t_2>1$ [see Fig.~\ref{fig.2}(a1) the regime $-\pi/2<\theta<-\pi/2$ and the regime outside that, respectively]. Apparently, the energy [see Eq.~(\ref{eq:refname4})] sensitively depends on the value of $\gamma$. Considering the expressions of $h_{x,y,z}$, the lattice can be found in three distinct phases: $\gamma<|t_1-t_2|$ for unbroken $\mathcal{PT}$ phase, $|t_1-t_2|<\gamma<|t_1+t_2|$ for partially broken $\mathcal{PT}$ phase, and $\gamma>|t_1+t_2|$ for fully broken $\mathcal{PT}$ phase. Here, the coupling parameters are given by $t_{1,2}= t (1 \mp \delta cos\theta)$, the parameter $\delta$ is the dimerization strength defined as $|\delta|$ < 1 and $\theta$ is an introducted tuning parameter varying from $-\pi$ to $\pi$, and $t$ can be viewed as an effective coupling constant. Besides, $\gamma_{c,\theta}=|t_1-t_2|=2 \delta t cos\theta$ is the critical value where $\mathcal{PT}$ phase transition occours.

For a $\mathcal{PT}$-symmetric system, the energy spectrum $E$ of the system under the open boundry conditions (OBC) ($N$=24) is investigated numerically as shown in Fig.~\ref{fig.2}. The qusi-1D photonic lattice has the topologically nontrivial phase in the regime of $-\pi/2<\theta<\pi/2$ [marked by the blue dashed lines $l$ and $l'$], characterized by the presence of near-zero edge states under the OBC, whereas no edge states exist in the regimes of $-\pi<\theta<-\pi/2$ and $\pi/2<\theta<\pi$ (or equally the regime of $-3\pi/2<\theta<-\pi/2$), corresponding to topological trivial phase [see Fig.~\ref{fig.2}(a1)]. It can be clearly seen that the eigenvalue spectrum of the system in regimes of $-\pi/2<\theta<\pi/2$ and $-3\pi/2<\theta<-\pi/2$ show different features for different $\gamma$. For the case of weak imaginary site potentials with $\gamma=0.1$, one can observe that midgap states with $Re(E)$ approaching 0 exist in the regime of $-\pi/2<\theta<\pi/2$ [see Fig.~\ref{fig.2}(a1)], and there are only two complex eigenvalues with a pair of conjugated imaginary parts in the same domain [see Fig.~\ref{fig.2}(a2)]. However, in the regime of $-3\pi/2<\theta<-\pi/2$, the non-Hermitian system has an entirely real eigenvalue spectrum, which indicates that the $\mathcal{PT}$ symmetry is unbroken in the presence of weak imaginary  potentials. When increasing the site potentials to $\gamma$=0.5 [see Figs.~\ref{fig.2}(b1) and (b2)], topological near-zero modes with two complex eigenvalues still exist in the regime of $-\pi/2<\theta<\pi/2$, it is worth noting that bulk states in two regimes begin to degenerate (the regimes between dashed lines $m(\theta = -\left| \arccos(\frac{\gamma_c}{2\delta t})\right|)$ and $l(\theta =-\pi/2)$,  $m'(\theta = \left| \arccos(\frac{\gamma_c}{2\delta t})\right|)$ and $l'(\theta =\pi/2)$, respectively) if $\gamma$ is larger than a critical value of $\gamma_{c,\theta}$, here $\gamma_{c,\theta}$ is the threshold where $\mathcal{PT}$ phases transition occurs. We know the $\gamma_{c,\theta}=|t_1-t_2|=2 \delta t cos\theta $. With $\gamma$ increasing, smaller regimes near $\theta=\pm\pi/2$ display complex eigenvalues corresponding to the breaking of $\mathcal{PT}$ symmetry. On the other hand, eigenvalues away from the regime $-\pi/2<\theta<\pi/2$ are entirely real, and thus the system still has unbroken $\mathcal{PT}$ symmetry. We find that the complex eigenvalues begin to emerge when $\gamma>\gamma_{c,\pm\pi}=1$. As shown in Fig.~\ref{fig.2}(c2), when $\gamma=1.1$, complex eigenvalues emerge in the whole regime of $\theta$, which indicates that the $\mathcal{PT}$ symmetry of systems in the whole regime is broken. 

\begin{figure}
\centering
\fbox{\includegraphics[width=\linewidth]{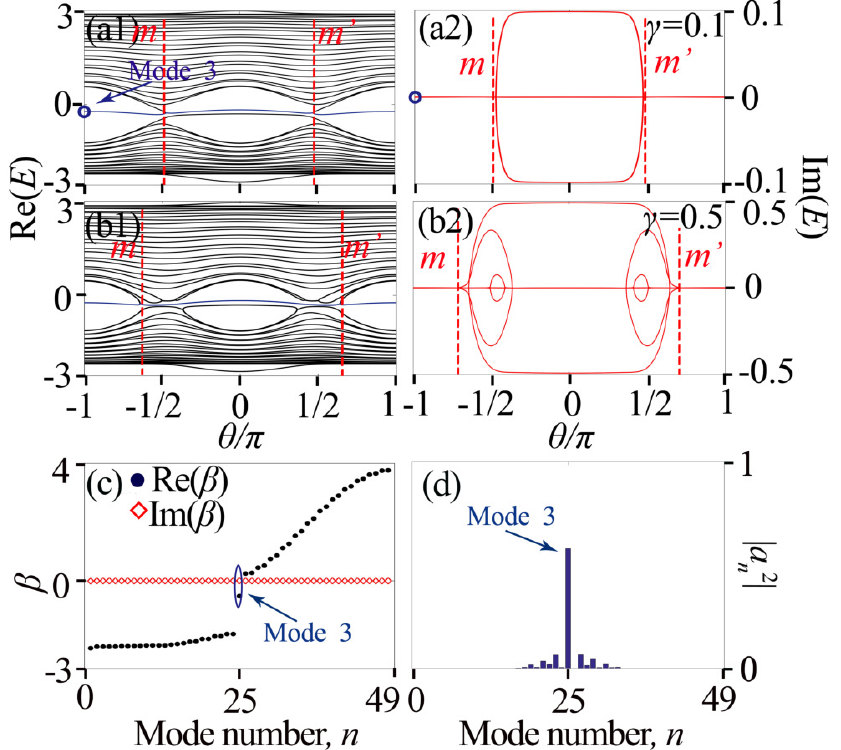}}
\caption{Eigenvalue spectrum and topological near-zero interface mode in $\mathcal{PT}$-symmetric binary photonic lattice in Fig.~\ref{fig.1}(b).  (c), (d), Propagation constants and intensity distributions of stable topological near-zero Mode 3 in Figs.~\ref{fig.3}(a1) and (a2). Here, \textit{t}=3/2, $\delta$=1/3, $\kappa$=0.2 and $\theta=\pm\pi$.}
\label{fig.3}
\end{figure}

Actually, the non-Hermitian term $\gamma$ can lead to different effects on the properties of the eigenvalue spectrum in topologically trivial and nontrivial phases. In the topologically trivial phase, the system undergos an abrupt transition from the unbroken $\mathcal{PT}$-symmetry region to the broken $\mathcal{PT}$-symmetry region at a certain $\gamma_c$. However, in the topologically nontrivial phase, the $\mathcal{PT}$ symmetry of the system is spontaneously broken once $\gamma$ is nonzero, which is characterized by the emergence of a pair of conjuguated complex near-zero modes. Consequently, topological edge states in such conventional $\mathcal{PT}$-symmetric system either on the sites with gain or on sites with loss, must have complex eigenvalues despite the fact that the $\mathcal{PT}$ operator still commutes with the Hamiltonian $\mathcal{H}$. Figure.~\ref{fig.2}(d) shows the eigenmodes of our quasi-1D $\mathcal{PT}$-symmetric topological lattice. It can be seen that all bulk states are in $\mathcal{PT}$-symmetric phase with real eigenvalues while the two topological near-zero edge mode circled in blue and cyan ellipses experience gain and loss, respectively. The corresponding amplified (Mode 1) and attenuated mode (Mode 2) intensities in Figs.~\ref{fig.2}(e) and (f) further confirm that topological edge states in such a conventional $\mathcal{PT}$-symmetric system either located on the site with gain or loss, must have complex eigenvalues despite the fact that the $\mathcal{PT}$ operator still commutes with the Hamiltonian $\mathcal{H}$.

To obtain a stable $\mathcal{PT}$-symmetric topological near-zero state, we proposed a binary quasi-1D structure with mirror-symmetric coupling constants and anti-mirror-symmetric gain and loss as well as an interface waveguide $a_D$ [see Fig.~\ref{fig.1}(b)]. The corresponding energy spectra and mode intensity are shown in Fig.~\ref{fig.3}. A midgap topological near-zero mode exists across the whole range of $-\pi<\theta<\pi$ for the introduction of the interface wavguide $a_D$. The topological near-zero modes in the regime $-\pi/2 < \theta < \pi/2$ are unstable due to the  spontaneously $\mathcal{PT}$ breaking arising from the gain and loss, which is similar to the analysation in Figs.~\ref{fig.2}(a1) and (a2); Thus, our focus turns to the regime of $-3\pi/2<\theta<-\pi/2$, the system undergoes an abrupt transition from the unbroken $\mathcal{PT}$-symmetry to the broken $\mathcal{PT}$-symmetry at a certain $\gamma_c$ in this regime. As shown in Figs.~\ref{fig.3}(b1) and (b2), the region between red dashed lines $m$ and $m'$ represent the $\mathcal{PT}$ broken system under a certain $\gamma$ and the midgap state with real eigenvalues for the interface waveguide ($a_D$) still exists outside the region[see Figs.~\ref{fig.3}(b1) and (b2)]. The exceptional points for the phase transition are $\theta_1$ and $\theta_2$ under a certain $\gamma$. This means stably topological near-zero interface state could be achieved once the gain/loss $\gamma$ is chosen precisley. 

Figure~\ref{fig.3}(c) and (d) show the calculated propagation constants and its intenstiy for Mode 3. The coupling constants is chosen to be $t_1>t_2$ in the regime of $-3\pi/2<\theta<-\pi/2$ to avoid the presence of edge modes on either side of the lattice. We can see the whole spectrum has exclusively real eigenvalues under the case of $\gamma$=0.1, and thus the system is $\mathcal{PT}$-symmetric. Therefore, such a binary quasi-1D $\mathcal{PT}$ system can be used to achieve the stably topological near-zero interface state.

\begin{figure}
\centering
\fbox{\includegraphics[width=\linewidth]{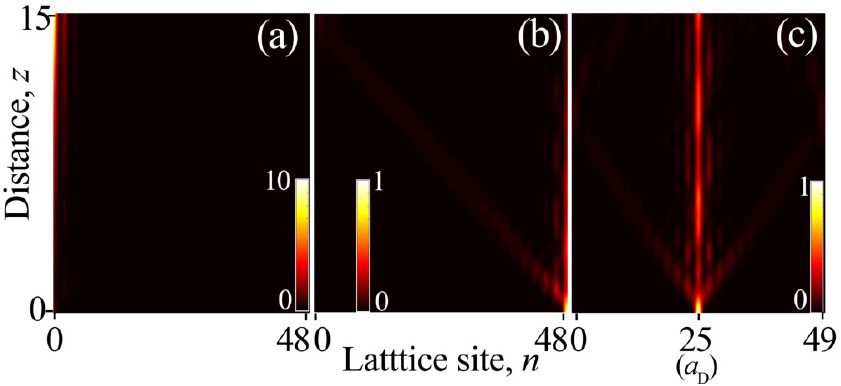}}
\caption{Light evolution in two different topological photonic lattice with $\mathcal{PT}$-symmetry in Fig.~\ref{fig.1}.  (a),(b) Topological edge states in Fig.\ref{fig.1}(a) experiencing gain and loss, respectively.  (c), Light evolution of after a single-waveguide excitation of the topological interface ($a_D$) in binary $\mathcal{PT}$-symmetric lattice. Here, $t$=3/2, $\delta$=1/3, $\theta=-\pi$ and $\gamma$=0.1.} 
\label{fig.4}
\end{figure}

Figure.~\ref{fig.4} shows the numerical simulations of the light propagation in our system with the site potential magnitude $\gamma=0.1$, corresponding to Mode 1, 2 and 3, respectively. When we launch a Gussian beam with a function of $v_0 \exp [- (n-n_0)^2/\omega_0^2 + i k_0 n]$ (here, $v_0 $, the light amplitude; $n_0 $, the waveguide number; $k_0$ , the wave vector; $\omega_0 = 0.5$, the normalized beam waist.) into the left-most waveguide of layer a ($a_1$), the light is localized and propagates along the left edge but experiences gain along the propagation [see Fig.~\ref{fig.4}(a)]. On the contrary, the light experience loss when it is injected from the right-most of the layer b ($b_{48}$) [see Fig.~\ref{fig.4}(b)], both of them behave unstable during their propagation. However, when we excite the interface waveguide ($a_D$=$a_{25}$) with the Gaussian beam, a stable topological interface mode with real eigenvalue can be realized in such a $\mathcal{PT}$-symmetric system [see Fig.~\ref{fig.4}(c)]. These results are in good agreement with the aforementioned theoretical and numerical studies, further confirming that stable  topological near-zero interface mode can be achieved in quasi-1D photonic lattice with $\mathcal{PT}$ symmetry. An applicable system with optical gain and loss might be performed in two-layered waveguide arrays based on Fe-doped LiNbO$_3$ crystal~\cite{R2010Observation}. The two-layered waveguide array is deeply modulated (see the red and blue waveguides in Fig.~\ref{fig.1}) where the losses arise from the optical excitation of electrons from Fe$^{2+}$ centers to the conduction band, and the optical gain can be provided through two-wave mixing using the material’s photorefractive nonlinearity~\cite{Kip1992Anisotropic}. 

In conclusion, we have studied the topological near-zero edge states systematically in a qusi-1D photonic lattice with $\mathcal{PT}$ symmetry and proposed a binary $\mathcal{PT}$-symmetric photonic lattice to realize a topological near-zero interface mode with real-eigenvalue. The analytical results show that the topological edge states are spontaneously broken in such a $\mathcal{PT}$-symmetric system and a stable near-zero interface state can be attained under certain conditions. Our work brings a novel way to obtain the $\mathcal{PT}$-symmetric near-zero mode in qusi-1D topological photonic lattice and can be used in future quantum computation.

\section{Funding Information}
This work is supported by the Open fund of Shandong Provincial Key Laboratory of Optics and Photonic Devices(K202008), Open Research Fund of State Key Laboratory of Transient Optics and Photonics (SKLST201805) and Natural Science Foundation of Shaanxi Province, China (2017JM6014).

\section{Disclosures}
The authors declare no conflicts of interest.

\bibliography{draft_ref}

\bibliographyfullrefs{draft_ref}


\ifthenelse{\equal{\journalref}{aop}}{%
\section*{Author Biographies}
\begingroup
\setlength\intextsep{0pt}
\begin{minipage}[t][6.3cm][t]{1.0\textwidth} 
  \begin{wrapfigure}{L}{0.25\textwidth}
    \includegraphics[width=0.25\textwidth]{john_smith.eps}
  \end{wrapfigure}
  \noindent
  {\bfseries John Smith} received his BSc (Mathematics) in 2000 from The University of Maryland. His research interests include lasers and optics.
\end{minipage}
\begin{minipage}{1.0\textwidth}
  \begin{wrapfigure}{L}{0.25\textwidth}
    \includegraphics[width=0.25\textwidth]{alice_smith.eps}
  \end{wrapfigure}
  \noindent
  {\bfseries Alice Smith} also received her BSc (Mathematics) in 2000 from The University of Maryland. Her research interests also include lasers and optics.
\end{minipage}
\endgroup
}{}

\end{CJK*}
\end{document}